\documentclass[english,aps,prstper,reprint,showpacs,titlepage,longbibliography]{revtex4-1}   

\usepackage[T1]{fontenc}	% should generally be included for better accented-word behavior
\usepackage[latin9]{inputenc}	% should generally be included for better accent behavior
\usepackage{geometry}		% for controlling page margins
\usepackage{graphicx}
\usepackage[above,below]{placeins}	% allows use of \FloatBarrier command to force section barriers
\usepackage{times}
\usepackage{color}

\usepackage{hyperref}  
\hypersetup{colorlinks=true,urlcolor=blue,citecolor=blue,linkcolor=blue}   
\usepackage{enumitem}          % package for handling list formatting
\setlist{nosep}                 % Tightest spacing for lists. `noitemsep` is more relaxed

\begin{document}

\begin{titlepage}

  \title{Faculty survey on upper-division thermal physics content coverage}

  \author{Katherine D. Rainey}
  \affiliation{Department of Physics, University of Colorado, 390 UCB, Boulder, CO 80309} 
  \author{Bethany R. Wilcox}
  \affiliation{Department of Physics, University of Colorado, 390 UCB, Boulder, CO 80309} 

  % \keywords{}

  \begin{abstract}
Thermal physics is a core course requirement for most physics degrees and encompasses both thermodynamics and statistical mechanics content. However, the primary content foci of thermal physics courses vary across universities. This variation can make creation of materials or assessment tools for thermal physics difficult. To determine the scope and content variability of thermal physics courses across institutions, we distributed a survey to over 140 institutions to determine content priorities from faculty and instructors who have taught upper-division thermodynamics and/or statistical mechanics. We present results from the survey, which highlight key similarities and differences in thermal physics content coverage across institutions. Though we see variations in content coverage, we found 9 key topical areas covered by all respondents in their upper-division thermal physics courses. We discuss implications of these findings for the development of instructional tools and assessments that are useful to the widest range of institutions and physics instructors.  \clearpage
  \end{abstract}
  %% Adding the `\clearpage` is the hack to make the title page.  In 2020, the proceedings is
  %% going to be double blind.  This change makes it so that we can programmatically remove the
  %% title page.  In the future, other blinding measures should be taken as well (for example,
  %% removing self-citations).  This is not needed in 2019.

  \maketitle
\end{titlepage}

% INTRO
\section{\label{sec:intro}Introduction}
Thermal physics, which includes both thermodynamics and statistical mechanics, is a core course required for attaining a physics bachelors' degree at most institutions. However, anecdotally the material covered in thermal physics courses often varies between instructors and across institutions. This content variability poses a significant challenge in development of standardized thermal physics assessments and teaching tools that can be utilized by a wide range of instructors. Though there is a body of research surrounding student understanding of thermal physics concepts \cite{dreyfus2015resource}, less is known about the breadth of topics covered in upper-division thermal physics courses. 

Here, we present findings from a survey distributed with the purpose of soliciting instructor priorities in upper-division thermal physics as a part of a broader research effort to develop a standardized upper-division thermal physics assessment. Findings may lay an important foundation for other researchers interested in developing course materials and assessments for thermal physics, and inform instructors in defining course objectives and content-foci for their thermal physics courses. 

In this paper, we begin by describing the process of constructing and distributing the survey (Sec. \ref{sec:methods}). Then, we present results of the survey (Sec. \ref{sec:results}), including general course information, key concepts covered, and valued scientific practices. We also consider response consistency between survey responses and submitted syllabi, followed by an analysis of content variability across institutions. We conclude with a short consideration of implications of the survey and future directions (Sec. \ref{sec:conclusion}).

% METHODS
\section{\label{sec:methods} Methods}

The faculty survey was designed to solicit key information about thermal physics courses, such as content covered, general course structure and emphasis (thermodynamics, statistical mechanics, or both), and needs or interest in an upper-division thermal physics assessment. This section describes methods for developing and distributing the survey with an emphasis on creating a format that was accessible and relatively short in duration, while still soliciting sufficient information.
%\vspace{-5mm}

\noindent{\bf Survey Development:} Prior to constructing the survey, a focus group was conducted with four experts, all with experience teaching thermal physics and researching student difficulties in thermal physics. The focus group solicited expert perspectives surrounding upper-division thermal physics, including textbooks, content coverage, learning goals, and existing thermodynamics assessments. Outcomes from the focus group informed several questions included on the survey. For example, participants discussed notational conventions as one major challenge for a thermal physics assessment (e.g. the sign convention of work). To address this concern, one question on the survey solicited specific notational issues worth considering in development of a thermal physics assessment. Additionally, textbooks brought up during the focus group comprised the list of textbook options provided on the survey.

To faciliate ease of responses, the survey was a primarily multiple-response format with only a select set of questions being free-response. Thus, one of the first steps in survey development was determining which options to provide for various multiple-response questions. We began by investigating the scope of thermal physics in texts; we analyzed six thermal physics texts brought up during the focus group \cite{baierlein1999thermal, carter2001csthermo, kittel1980thermal, salinger1975thermo, schroeder1999thermal, zemansky10020heat} for key content coverage. This process involved reviewing each text and identifying topical areas for each based on chapter titles, section headings, and emphasized key terms. Based on the frequency of topics appearing across the different texts, we classified topical areas into \emph{core topics} and \emph{supporting topics}. To put these into an accessible form for use in the survey, topics were sorted and condensed into 29 core topics, most with roughly 4 supporting topics (see Table \ref{table:contentfreq}). For example, the core topic of ``thermodynamic laws'' had four supporting topics: 0th law, 1st law, 2nd law, 3rd law. Some core topics had no supporting topics (e.g. semiconductors) while some had as many as seven (e.g. energy and thermodynamic potentials); the one exception to this was statistical mechanics, which had 14 supporting topics.

In addition to focusing on content, and in response to recent calls in science education literature for more consideration of scientific practices in course materials, assessment, and instruction \cite{national2012framework}, the survey also solicited information on the scientific practices valued by respondents in their thermal physics courses. The list of scientific practices provided on the survey was pulled from the Next Generation Science Standards (NGSS) list of science and engineering practices \cite{NGSS2013practices}. In their list, the NGSS combined similar practices together (e.g. developing and using models); however, in upper-division courses, it is less clear that all paired practices would be targeted together. Thus, to collect more specific data about individual practices, paired NGSS practices were split into separate categories. For example, ``developing and using models'' was split into ``developing models'' and ``using models'' for the survey. 

The survey was administered through the survey platform Qualtrics and hosted by the University of Colorado Boulder (CU). The survey was divided into 4 major sections: (1) general course information, (2) content coverage, (3) scientific practices, and (4) interest in, and concerns about, an upper-division thermal physics assessment. Respondents also had the option to identify their institution and submit their course syllabus. Additionally, gender and racial identity information were collected at the end of the survey. 

After initial construction of the survey, we solicited feedback from CU physics faculty who were familiar with teaching upper-division thermal physics. Based on these discussions, and informed by the frequency of topical areas appearing across the six different analyzed texts, we grouped the core topics into two categories: assumed core topics and other core topics. Assumed core topics are topics that one might expect are covered in every thermal physics course: energy and thermodynamic potentials; engines and refrigerators; entropy; equilibrium; monatomic gases; heat; temperature; thermodynamic laws; and work. The survey presented these assumed core topics at the beginning of Section (2) of the survey, with their supporting topics shown on the same page. A free-response textbox followed these assumptions to allow respondents to indicate disagreement with the assumptions made. All other core topics were provided on the following page of the survey without their supporting topics displayed. After selecting from the list of other core topics, associated supporting topics for each of the selected core topics were displayed on the following page. This conditional formatting was motivated by the desire to reduce respondent fatigue due to survey length. 

\noindent{\bf Survey Distribution:} To ensure the information collected was reflective of a broad range of institutions, we collected contact information for a large variety of physics degree-granting institutions, including minority serving institutions (MSIs) and women's colleges, for use in distributing the survey. Institutions were identified using the American Physical Society's ``Top Educators'' lists \cite{APSTopEd}, each of which identifies 16-20 institutions with the highest average number of physics bachelors' degrees awarded by the institution per year. We also utilized the overall and underrepresented minority (URM) lists for Ph.D.-granting, MS-granting, and BS-granting institutions. Beyond that, we used the American Physical Society's MSIs list \cite{MSI2017}, which included a list of Historically Black Colleges and Universities, Black-serving insitutions, and Hispanic-serving institutions, to identify all other physics-degree-granting MSIs not on the ``Top Educators'' lists; the MSI list included institutions with both large and small physics departments. We also identified women's colleges with the ``Women in Physics'' report produced by the American Institute of Physics \cite{women2000AIP}. We note that other small physics departments (e.g. those that are not Top Educators, or at MSIs or women's colleges) were not targeted in the initial distribution of the survey, but will be targeted in the broader project moving forward. 

After identifying institutions, we obtained contact information of department chairs from physics department websites. We then emailed the survey solicitation to the department chairs, with a specific request for the email to be forwarded to all faculty within their department who were currently teaching or had previously taught upper-division thermal physics. In addition to department chairs, the research team solicited the help of their professional contacts at different institutions to take the survey or forward it to faculty in their department.  

% RESULTS
\section{\label{sec:results}Results}

The survey was open for response collection for three and a half months. During this time, 59 respondents fully completed the survey while 2 completed all of the survey except questions regarding scientific practices and assessment. Only responses that completed the sections with core topics and supporting topics and beyond were used for analysis. We do not report response rate, as it is unclear how many people recieved the solicitation forwarded from their department chairs. 

\begin{figure}
  \includegraphics[width=\linewidth]{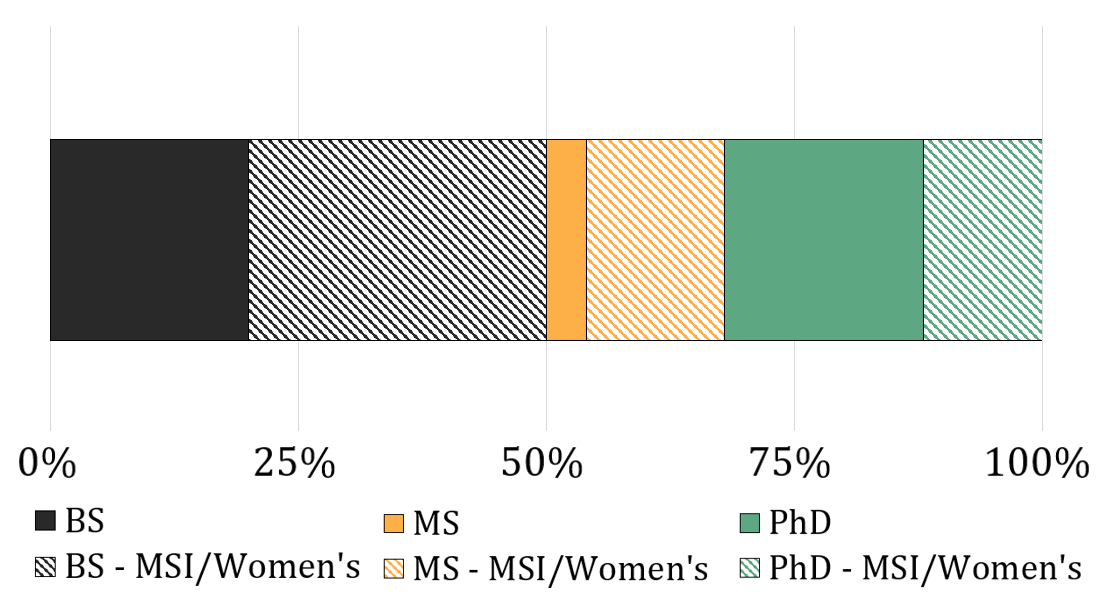}
  \caption{Highest physics degree offered by Minority-Serving Institution (MSI) or Women's College classification. Bachelor's degrees (BS), Master's degrees (MS), and PhDs are indicated. \label{MSI}}
\end{figure}

Racial demographics of respondents included Asian (16\%, N=9), Black/African American (2\%, N=1), Caucasian (74\%, N=43), and Hispanic (2\%, N=1); no other racial identities were indicated and 7\% (N=4) preferred not to answer. Additionally, 83\% (N=48) of respondents were men and 14\% (N=8) were women (no other gender identities were indicated); 3\% (N=2) preferred not to provide their gender. Three respondents did not provide any demographic information.

\begingroup
\begin{table*}
\caption{Response frequency for thermal physics content. The two left columns show data for assumed topics. All assumed core topics appeared at a frequency of 100\%. The right column shows all other core topics. No supporting topics are presented for other core topics. Topics that appeared on syllabi but not the survey (e.g. ensembles and thermodynamic identities) are also not presented.}

	\begin{ruledtabular}
		\begin{tabular}{l c l c l c}

		\multicolumn{1}{c}{Assumed Topic} & \multicolumn{1}{c}{\%} & \multicolumn{1}{c}{Assumed Topic} & \multicolumn{1}{c}{\%}& \multicolumn{1}{c}{Other Core Topic} & \multicolumn{1}{c}{\%} \\ 
		\hline \hline 
		Energy \& Thermodynamic Potentials & & Equilibrium & & Statistical Mechanics & 92 \\ 
		\hspace{2mm} \emph{Chemical Potential} &  93  & \hspace{2mm} \emph{Thermal Equilibrium} & 98  & Processes & 89 \\
		\hspace{2mm} \emph{Energy Sources} &  49  & \hspace{2mm} \emph{Stable \& Unstable Equilibrium} & 41 & Diatomic Gases & 84 \\
		\hspace{2mm} \emph{Enthalpy} &  89  & Heat &  & Fermions & 84 \\
		\hspace{2mm} \emph{Equipartition} &  95  & \hspace{2mm} \emph{Heat Capacity} & 100  & Blackbody Radiation & 82 \\
		\hspace{2mm} \emph{Free Energy (Gibbs \& Helmholtz)} & 95 & \hspace{2mm} \emph{Heat Transfer} & 72 & Bosons & 80 \\
		\hspace{2mm} \emph{Internal Energy} &  100  & \hspace{2mm} \emph{Latent Heat} & 90  & Phases & 79 \\
		\hspace{2mm} \emph{Maxwell's Relations} &  77  & Temperature &  & Kinetic Theory & 75 \\
		Engines \& Refrigerators &   & \hspace{2mm} \emph{Absolute Zero}  & 98 & Quantum Phenomenon & 75 \\
		\hspace{2mm} \emph{Heat Engines} & 93 & \hspace{2mm} \emph{Negative Temperature} & 69 & Pressure Diagrams & 72 \\
		\hspace{2mm} \emph{Refrigerators} &  82  &  \hspace{2mm} \emph{Thermodynamic Temperature} & 89 & Scaling & 71 \\
		Entropy &  & \hspace{2mm} \emph{Temperature Measurement} & 59 & Magnetism & 64 \\
		\hspace{2mm} \emph{Boltzmann's Law} &  90  & Thermodynamic Laws & & Chemical Reactions & 54 \\
		\hspace{2mm} \emph{dS=dQ/T} & 93 & \hspace{2mm} \emph{0th} & 89  & Conduction, Convection, Radiation & 53 \\
		\hspace{2mm} \emph{Entropy \& Information} &  57  & \hspace{2mm} \emph{1st} & 100 & Solids & 51 \\
		\hspace{2mm} \emph{TS Diagrams} &  71  & \hspace{2mm} \emph{2nd} & 100  & Pure Substances & 49 \\
		Gases &   & \hspace{2mm} \emph{3rd} & 89 & Diffusion & 46 \\
		\hspace{2mm} \emph{Ideal Gas Law} &  100  & Work &  & Cooling Techniques & 31 \\
		\hspace{2mm} \emph{Mixtures of Gases} & 57  & \hspace{2mm} \emph{Mechanical} & 98 & Fluids & 20 \\
		\hspace{2mm} \emph{van der Waals Interactions} &  71  & \hspace{2mm} \emph{Path dependence} & 84 & Semiconductors & 12 \\

		\end{tabular}
	\end{ruledtabular} \vspace{-.5\baselineskip}

\label{table:contentfreq}
\end{table*}
\endgroup

We collected institutional information, including selectivity, research activity, student population, and highest physics degree offered via the Carnegie Classifications \cite{Carnegie2018} and institutions' physics department websites. From the Carnegie Classifications, we identified 70\% (N=34) of identifiable institutions as being selective or more selective with regards to admissions practices, while 31\% (N=15) are considered ``inclusive'' institutions. Additionally, 18 schools are classified as having high or very high research activity.  

Overall, we identified 52 unique institutions from the survey, 28 of which were MSIs and/or women's colleges; one institution could not be identified and one was not in the Carnegie Classifications database. Figure \ref{MSI} presents institution type by highest physics degree offered and MSI/women's college classification. In a few cases, (N=7) institutions were represented by 2-3 responses; it was evident from submitted syllabi and individual item reponses that these were submitted by different people. 

\noindent{\bf Course Information:} We asked respondents if their course focused on thermodynamics, statistical mechanics, or both (thermal physics); 97\% (N=59) selected thermal physics and the remaining 3\% (N=2) of responses were split evenly between thermodynamics and statistical mechanics. Most institutions reported one semester of thermal physics (79\%, N=48); some reported two quarters (10\%, N=6) or two semesters (8\%, N=5), while a small minority reported one quarter (3\%, N=2). The student population was composed of mostly juniors (N=41) and seniors (N=39), though some (N=12) reported sophomores in the course as well. 

The majority of respondents (72\%, N=44) reported using \textit{An Introduction to Thermal Physics} by Daniel V. Schroeder \cite{schroeder1999thermal}. \textit{Thermal Physics} by Charles Kittel and Herbert Kroemer \cite{kittel1980thermal} was the second most frequently cited text (16\%, N=10). All other texts appeared at a frequency of 7\% or below. Most of the instructors (74\%, N=45) teach with the assumption that their students have little to no prior exposure to thermal physics content. Some (N=19) expected familiarity with topics such as energy, heat, the first and second laws of thermodynamics, and the ideal gas law. A few (N=7) said they expect thermal physics exposure from the introductory physics sequence, though several noted that thermal physics is only covered for a few weeks, and sometimes not at all, in that sequence.

These data show most institutions require one semester of thermal physics, most instructors use Schroeder's text \cite{schroeder1999thermal}, and many instructors assume their students have no prior exposure to thermal physics content. These results suggest two implications for PER: (1) development of Schroeder-based thermal physics assessments and materials could serve many instructors and institutions, though would still exclude the sizable population of instructors and institutions who do not use that text; and (2) pretest administration of an upper-division thermal physics assessment may not produce meaningful measurements of student understanding of thermal physics content prior to taking the course.

\noindent{\bf Key Topical Areas:} Table \ref{table:contentfreq} shows frequency of assumed supporting topics and other core topics. All assumed core topics (see Section \ref{sec:methods}) appeared at a frequency of 100\%; these frequencies are not reported in Table \ref{table:contentfreq}. Frequency of supporting topics is given relative to the number of times the corresponding core topic was selected; the frequency of core topics is given relative to the total number of valid responses. We present frequencies of all other core topics, but do not present their 56 associated supporting topics or their frequencies due to space limitations. 

Four respondents reported teaching thermal physics but did not select statistical mechanics as a core topic. This result may be due to statistical mechanics being covered in their course but not seen as a core focus by the respondent; we note one of these respondents mentioned statistical distribution functions in a textbox but did not select statistical mechanics as a core topic. 

These results are relevant for researchers interested in materials and assessment development in upper-division thermal physics, and can be used to guide content-foci for those endeavors such that they serve a wide range of instructors and institutions. 
\newline

\noindent{\bf Scientific Practices:} Of the 16 practices presented on the survey, three appeared at a frequency of over 85\%: using mathematical thinking (98\%, N=58), asking questions (95\%, N=56), and using models (86\%, N=51). Review of syllabi indicates the practice of ``asking questions'' may have been misinterpreted; the NGSS practice refers to asking scientific questions (namely for scientific investigations), but we suspect respondents may have interpreted this practice as referring to asking questions about content during class or office hours. The next most frequently appearing practices were constructing explanations (70\%, N=41), communicating information (64\%, N=38), and computational thinking (61\%, N=36). The remaining 10 practices appeared at a frequency of 56\% or less. 

These results highlight at most three scientific practices that stand out as valued by nearly all thermal physics instructors in our sample and demonstrate many other scientific practices are less of a universal focus for thermal physics courses at the upper-division level. Thus, researchers should pay particular attention to including opportunities for students to demonstrate and develop the practices of using models and using mathematical thinking in thermal physics-oriented materials and assessments. 

\noindent{\bf Response Consistency:} As a verification of the survey data, we checked for consistency between survey responses and submitted syllabi for the 39 responses that provided a syllabus. We looked at key topics on syllabi and compared with the associated survey response to ensure topics appearing on the syllabus also appeared on the survey response. No core or supporting topics had more than 3 discrepancies when comparing between survey responses and the 39 syllabi. Discrepancies could be due to the amount of focus placed on those topics in the course. For example, Bose-Einstein condensates may appear on the syllabus but may not be seen as a major content focus for the instructor when completing the survey, resulting in a discrepancy between their syllabus and response. Some topics, such as large systems (N=10), interacting systems (N=8), and Boltzmann and/or quantum statistics (N=9), appeared in syllabi but did not appear as explicitly named core or supporting topics on the survey. However, those who included topics such as these on their syllabus selected other topics on the survey that encompass or require the same idea, such as multiplicity, thermal equilibrium, and statistical mechanics. Canonical ensembles (N=11) and thermodynamic identities (N=6) were the other most common topics that appeared on syllabi but were not provided as options on the survey.

This analysis shows that the survey reliably captured the scope of content coverage for most survey responses without large discrepancies.

\noindent{\bf Content Variability:} To investigate the claim of content variation across upper-division thermal physics courses, we examined survey responses to see how many topics were selected by all instructors. We looked at the three groups of topics laid out in Table \ref{table:contentfreq}: assumed core topics, assumed core topics' supporting topics, and other core topics.

We found that 9/9 (100\%) of assumed core topics, 5/32 (16\%) of assumed supporting topics, and 0/20 (0\%) of other core topics were selected by all respondents. When repeated with institutions with multiple responses (e.g. different instructors at the same institution), we saw an average of 72\% of assumed supporting topics and 20\% of other core topics chosen by all respondents at a given institution. 

%and determined the fractional amount of topics within those groups that all respondents selected. This analysis yeilds a scale in which 0.00 indicates no alignment across the topic group and 1.00 represents complete alignment across the considered topic group. For example, if a topic group contained 10 topics where 6 of those topics were selected by all respondents, this would correspond with an alignment of 0.60. {\color{red}{This would mean that 0.60 of topics within that topic group appeared at a frequency of 100\%.}}

%As predicted, alignment for the group of assumed core topics was 1.00 {\color{red}{(9/9)}}. We considered two additional topic groups: assumed core topics' supporting topics {\color{red}{(N=32)}} and other core topics (excluding supporting topics) {\color{red}{(N=20)}}. Across all assumed supporting topics, there was 0.16 {\color{red}{(5/32)}} alignment between all institutions; this number was 0.00 {\color{red}{(0/20)}} for all other core topics. When repeated with institutions with multiple responses (e.g. different instructors at the same institution), we saw an average alignment of 0.72 for assumed supporting topics and 0.20 for other core topics. 

These results support the anecdotal claim that upper-division thermal physics content coverage varies both across institutions and between instructors at the same institution (though to a lesser extent). It also makes the case, however, that there are some topics, namely our assumed core topics, that all or most instructors prioritize in their upper-division thermal physics courses. 

% CONCLUSION
\section{\label{sec:conclusion}Conclusions}

Our data suggest important considerations for researchers and instructors interested in curricular materials and assessment development for upper-division thermal physics. Despite the demonstrated content variability within thermal physics, our results point to content-foci, scientific practices, and reference texts that can act as baselines for materials that can serve a broad range of institutions and instructors. The results presented here will lay the groundwork for development of an upper-division thermal physics assessment. In order for this assessment to be useful broadly, we carefully and deliberately collected data from institutions that serve a wide range of student populations. We recommend other researchers interested in making widely-available upper-division materials utilize similar methods in collecting input from a wide range of institutions to inform their work. Results from this survey can inform upper-division thermal physics investigations in PER and the methodology can be reproduced for investigation of the scope of other upper-division physics courses. 

% ACKNOWLEDGEMENTS
\acknowledgments{This work was supported by funding from the Center for STEM Learning and the Department of Physics at CU. We thank S. Pollock and M. Dubson for their input in refining the survey, J. T. Laverty for his encouragement of including scientific practices, and all focus group participants. We also thank the department chairs who distributed the survey and the faculty who completed it. We are grateful for their time.} 

\newpage
% REFERENCES
\bibliography{2019_PERC-refs}

\end{document}